\documentclass[aps,pre,10pt,twocolumn,superscriptaddress,showpacs,floatfix,longbibliography]{revtex4-1}
\usepackage{amsmath}
\usepackage{amssymb}
\usepackage{graphicx}
\usepackage{esint}
\usepackage[svgnames]{xcolor}
\definecolor{Darkred}{rgb}{0.8,0,0}
   \usepackage[colorlinks,bookmarks=false,citecolor=blue,linkcolor=Darkred,hyperfootnotes=true,urlcolor=blue,breaklinks]{hyperref}
\usepackage{url}

\begin{document}

\title{Quasi-local charges and the Generalized Gibbs Ensemble in the Lieb-Liniger model}

\author{T. Palmai}
\affiliation{Condensed Matter Physics \& Materials Science Division, Brookhaven National Laboratory, Upton, NY 11973-5000, USA}

\author{R. M. Konik}
\affiliation{Condensed Matter Physics \& Materials Science Division, Brookhaven National Laboratory, Upton, NY 11973-5000, USA}

\begin{abstract}
We consider the construction of a generalized Gibbs ensemble composed of 
complete bases of conserved charges in the repulsive Lieb-Liniger model.  We will show that it is possible to construct these bases
with varying locality as well as demonstrating
that such constructions are always possible provided one has in hand at least one complete basis set
of charges.  This procedure enables the construction of bases of charges that possess well defined, finite expectation values
given an arbitrary initial state.  We demonstrate the use of these
charges in the context of two different quantum quenches:  a quench
where the strength of the interactions in a one-dimensional
gas is switched suddenly from zero to some finite value
and the release of a one dimensional cold atomic gas from a confining parabolic trap.
While we focus on the Lieb-Liniger model in this paper, the principle of the construction of these charges applies to all integrable models, both in continuum and
lattice form. 
\end{abstract}

\maketitle
\section{Introduction}

It is widely accepted that if one pumps energy into a closed quantum system that relaxation to a steady state is governed by the presence
of all of the conserved quantities in the system, provided the system is in the thermodynamic limit \cite{Polkovnikov2011,Gogolin2016}.  If the conserved quantities or charges are labelled $\{\hat Q_i\}^K_{i=1}$,
where $K$ may be either finite or infinite, then the steady state reached by the system should be governed by a density matrix, $\hat\rho$
\begin{equation} \label{eI}
\hat\rho_\text{GGE} =\frac{1}{Z} e^{-\sum_{i=1}^K \beta_i \hat Q_i}.
\end{equation}
Here $\beta_i$ are the (generalized) temperatures associated with each charge $\hat Q_i$.  In this general way of describing relaxation
in a closed quantum system, two cases are usually separated out: 
i) one where there is only an intensive number of conserved quantities, perhaps only the Hamiltonian of the system itself
\cite{Deutsch1991,Srednicki1994,Srednicki1996,Srednicki1999,Rigol2008,Moeckel2008,Rigol2010,Banuls2011,Rigol2014}; and 
ii) one where there are an infinite set of conserved quantities that govern relaxation in the long time limit \cite{Rigol2007,Rossini2009,Rossini2010,Calabrese2011,Calabrese2012,Calabrese2012a,Essler2012,Mossel2010,Mossel2012,Iucci2009,Dora2012,Fioretto2010,Sotiriadis2012,Mussardo2013,Caux2012,Fagotti2014,Pozsgay2013,Ilievski2015,Essler2016,Vidmar2016,Kormos2014,Biroli2010,Gogolin2011,Cazalilla2012,Foini2012,Cassidy2011,Pozsgay2011,Sotiriadis2014,Goldstein2014,Pozsgay2014,Essler2015,Piroli2016,Bastianello2017,Kormos2013,Langen2015}.  In the first case the system is
said to relax to a standard Gibbsian ensemble governed by a single effective temperature while in the second case the system is said to be integrable and relaxation is instead to
a {\it generalized} Gibbsian ensemble (GGE.)

In the past few years a complementary view of relaxation in a closed {\it integrable} quantum system has arisen 
\cite{Caux2013,Caux2016,DeNardis2014,Wouters2014,Brockmann2014,Pozsgay2014,Mestyan2015,2017arXiv170701073B,SciPostPhys.1.1.001,Mestyan2017,Piroli2016,1742-5468-2016-6-063102} in response to difficulties in defining $\hat\rho_\text{GGE}$ in certain instances.  Rather than thinking of the long time behavior of the system being
governed by a density matrix involving the system's conserved quantities, i.e. Eqn. \ref{eI}, the notion of a `representative state' is employed.  Whereas the density
matrix of Eqn. \ref{eI} is associated with a canonical ensemble, a representative state is invoked by combining a (generalized) microcanonical ensemble with
the (generalized) eigenstate thermalization hypothesis \cite{Deutsch1991,Srednicki1994,Srednicki1996,Srednicki1999,Pozsgay2011,DAlessio2016}.  For a generalized microcanonical ensemble, the density matrix reads
\begin{equation}
\hat\rho_{\text{mc},\{Q_i\}} = \sum_{|s\rangle, \langle  s| \hat Q_i|s \rangle \in \{Q_i -\epsilon , Q_i + \epsilon\}}|s\rangle\langle s|
\end{equation}
Here the density matrix is a sum of projection operators over all states $|s\rangle$ whose quantum numbers $\langle s| \hat Q_i | s\rangle$ fall
in a narrow range about the values $Q_i$.  What the generalized eigenstate thermalization hypothesis (gETH) argues is that the states $|s\rangle$
are all equally good for determining the long time properties of a system.  Specifically, for any reasonable observable ${\cal O}$, the gETH
states that for any state $|s\rangle$ involved in the sum of states composing the microcanonical ensemble we have
\begin{equation}
\langle s| {\cal O}| s \rangle \equiv {\rm Tr} \,\hat\rho_{\text{mc},\{Q_i\}}
\end{equation}
Thus the gETH reduces the problem of finding the longtime limit of an observable to computing a single expectation value.

Even with this view, there remains the problem of determining a representative state $|s\rangle$.  However here we have a number of
options.  Most generally, we have the quench
action \cite{Caux2013,Caux2016}.  The quench action defines a generalized action whose saddle point defines the representative state $|s\rangle$.  Finding the representative
state using the quench action has now been demonstrated in a number of instances: i) quenches in the transverse field Ising model \cite{Caux2013}, ii) quenches in the Lieb-Liniger model \cite{DeNardis2014,SciPostPhys.1.1.001}, iii) the Neel-to-XXZ \cite{Wouters2014,Brockmann2014,2017arXiv170510765A} and Majumdar-Ghosh (dimer)-to-XXZ \cite{Pozsgay2014,Mestyan2015} quenches in the XXZ Heisenberg spin chain,
iv) quenches in the Hubbard model \cite{2017arXiv170701073B}, v) quenches in spin-1 chains \cite{Mestyan2017,Piroli2016a}, and vi) quenches in relativistic field theories
\cite{1742-5468-2016-6-063102}.  Separate from the quench action for determining the representative state, we have, in the particular case of the XXZ model (and similar
integrable lattice models), the ability to relate the expectation values of a certain class of charges to the densities of excitations that characterize the representative 
state \cite{Ilievski2015,charge-string}.

One virtue that the quench action has is that it leads to physical results: the representative state, $|s\rangle$, that is determined as the saddle point of the quench action
has well-defined expectation values on local
observables. This need not be the case for a generalized Gibbs density matrix.  In particular, it need not be the case that the states $|s\rangle$ have
a finite expectation on the conserved quantities themselves, it may be that we have
\begin{equation}
\langle s| \hat Q_i |s\rangle = \infty,
\end{equation}
leading to difficulties in sensibly defining $\hat\rho_\text{GGE}$.

How this can happen is readily seen.  Typically if a system possesses an infinite set of conserved quantities beyond the Hamiltonian itself, these additional
conserved charges are often constructed by looking for, roughly speaking, higher moments of the Hamiltonian or energy-momentum tensor.  
And then, while the energy density, $E_s$, of a state
$|s\rangle$ may be finite, higher moments of the energy may diverge.  For example if a state has degrees of freedom each with energy $E$ and distributed
according to $\rho_s (E)$, the energy of the state can be written as
\begin{equation}
E_s = \int dE\, E \rho_s(E).
\end{equation}
And while the above integral may be convergent, the integral
\begin{equation}\label{conservedgen}
E^n_s = \int dE\, E^n \rho_s(E)
\end{equation}
determining a higher (n-th) moment of the energy may not be.  In this sense the quench action and its attendant representative states
have a certain practical advantage over the GGE density matrix -- it does not require that the conserved quantities have well defined expectation values.
It was this advantage that allowed the interaction quench in the Lieb-Liniger model to be fully described \cite{DeNardis2014}.

As we have said, the origin of this problem lies in the nature of the typical construction of the infinite hierarchy of conserved quantities
in an integrable model -- namely as higher moments of the energy density.  We show here that in fact that one is never limited to this
particular hierarchy and that in fact it is possible to construct conserved quantities which have generically finite expectation values.
We show that if there exists one complete basis of conserved charges (in a sense to be described), we can construct arbitrary bases of
charges.  We can always design these bases so that they are quasi-local, i.e. a quasi-local charge is a charge defined 
as an integral over space,
\begin{equation}
\hat Q = \int dx\, \hat q(x),
\end{equation}
where $\hat q (x)$ is an operator whose support is found primarily about the spatial position $x$.  We thus show that it is always
possible to have a well-defined a GGE for a given quench.

This builds on prior work on quasi-local charges in the quantum Ising field theory.  In Refs. \cite{Essler2015,Essler2017} it was shown that one can construct explicit 
quasi-local charges in the quantum Ising field theory.  The construction of the charges was possible because the underlying description of the 
model is that of free fermions.  Here we show that this construction can be generalized to arbitrary interacting theories.

We will demonstrate this construction in the context of the Lieb-Liniger model \cite{Lieb1963,Lieb1963a}.  This model offers several advantages here.  It is generically
interacting and so demonstrates the possibility of construction of alternative hierarchies of conserved quantities in interacting theories.
However it is also relatively simple, for example, in its repulsive regime it does not possess string solutions of its attendant Bethe Ansatz equations.
Moreover it has a limit where it maps onto free fermions -- which we
will exploit at times.

While our focus here will be on quasi-local charges in continuum theories, it would be remiss not to mention
that there has been considerable recent interest in quasi-local charges in lattice models \cite{Ilievski2015a,Ilievski2015,Ilievski_review,Ilievski2017,Pozsgay2017}.
Such quasi-local charges, constructed in the framework of the algebraic Bethe ansatz \cite{Ilievski2015a}, 
have been shown to be a necessary ingredient for GGEs describing the N\'{e}el quench in the XXZ Heisenberg model \cite{Ilievski2015}. 

The paper is organized as follows.  In Section 2 we provide an overview of the integrable structure of the Lieb-Liniger model.  In Section 3 we 
demonstrate how construction of arbitrary bases of conserved quantities is possible.  In Section 4 we construct explicit operatorial expressions for large but finite c
for the charges and show under what conditions the charges are quasi-local.
In Section 5 we apply these ideas to the interaction quench in the Lieb-Liniger model where the ultra-local charges fail to provide a sensible GGE, while in 
Section 6 trap-release quench is studied where GGEs based on both the ultra-local and the quasi-local charges can be sensibly defined.
Finally in Section 7 we wrap up with a discussion in the context of recent proposals for other alternatives to using ultra-local charges.

\section{Lieb-Liniger Model}
In this section we provide an overview of the integrable structure of the Lieb-Liniger model.
The Lieb-Liniger model describes a system of $N$ identical bosons on a one-dimensional ring of circumference $L$, interacting through a contact potential \cite{Lieb1963,Lieb1963a},
\begin{equation}\label{LLHam}
H=-\sum_{i=1}^N \frac{\partial^2}{\partial x_i^2}+2c\sum_{i<j}\delta(x_i-x_j),
\end{equation}
or in the second quantized form,
\begin{equation}
H=\int_0^L dx\left(-\Phi^\dagger(x)\partial^2_x\Phi(x)+c\Phi^\dagger(x)\Phi^\dagger(x)\Phi(x)\Phi(x)\right),
\end{equation}
where we set $\hbar=2m=1$ and $c$ is the interaction strength. We will work in the repulsive regime, $c>0$.

The exact eigenstates of (\ref{LLHam}) are described by the Bethe Ansatz wave function \cite{Korepin1993},
\begin{multline}
\xi(x_1,\ldots,x_N;I_1,\ldots,I_N)=\\
F_{\{I_j\}}\sum_\mathcal{P}
\left[\prod_{j>k}
\left(1-\frac{ic\,\text{sgn}(x_j-x_k)}{\lambda_{\mathcal{P}_j}-\lambda_{\mathcal{P}_k}}\right)
\right]
e^{i\sum_n x_n\lambda_{\mathcal{P}_n}},
\end{multline}
where $F_{\{I_j\}}=\frac{\prod_{j>k=1}^N(\lambda_j-\lambda_k)}{N!\prod_{j>k=1}^N((\lambda_j-\lambda_k)^2+c^2)}$, $\mathcal{P}$ is a list of all permutation of the indices and the quasi-momenta, 
$\lambda_n$, are determined by the Bethe equations \cite{Lieb1963,Lieb1963a} in terms of a set of $N$ distinct integers (half-odd integers) $\{I_j\}$ for $N$ odd (even),
\begin{equation}\label{bethe}
\lambda_j=\frac{2\pi I_j}{L}-\frac{1}{L}\sum_k\theta (\lambda_j-\lambda_k),\quad j=1,2\ldots,N
\end{equation}
and where the scattering phase $\theta(\lambda)$ equals
$$
\theta(\lambda) = 2\tan^{-1}\left(\frac{\lambda}{c}\right).
$$

When the thermodynamic limit (TDL) is approached, $L,N\to\infty$, and the particle density $n=N/L$ remains finite, the occupied 
quasimomenta or roots become continuous in $\lambda$ and it is useful to introduce a density function,
\begin{equation}
\rho_p(\lambda_j)=\frac{1}{L(\lambda_{j+1}-\lambda_j)}.
\end{equation}
In the TDL, the Bethe equations combine into
\begin{equation}\label{rho_eqn}
\rho_h(\lambda)+\rho_p(\lambda)=\frac{1}{2\pi}+\int\frac{d\lambda'}{2\pi}K(\lambda-\lambda')\rho_p(\lambda'),
\end{equation}
where we introduced the density of empty quasi momentum modes $\rho_h$ and the kernel 
$$
K(\lambda)\equiv \theta'(\lambda)=\frac{2c}{\lambda^2+c^2} .
$$
In the framework of the quench action, $\rho_p$ is a key quantity.  A given representative state, $|s\rangle$, is described
by specifying the distribution $\rho_{p,s}(\lambda)$ of particles in the state.

In the TDL the expression above for the rapidity (Eqn. \ref{bethe}) can be rewritten as 
\begin{equation}\label{bethe1}
\lambda_j=\frac{2\pi I_j}{L} + \int^\infty_{-\infty}d\lambda'F(\lambda',\lambda_j) n(\lambda').
\end{equation}
Here $n(\lambda)$ is a function bounded by $0$ and $1$ and is given in terms of $\rho_{p/h}(\lambda)$, 
the density of states for particles/holes at $\lambda$ via
\begin{equation}\label{F}
n(\lambda ) = \frac{\rho_p(\lambda)}{\rho_p(\lambda) + \rho_h (\lambda)}.
\end{equation}
The expression in the continuum limit for $\lambda_j$ furthermore involves the shift function $F(\lambda,\lambda')$:
\begin{eqnarray}
F(\lambda,\lambda') &=& \theta(\lambda-\lambda') \cr\cr
&& + \int^\infty_{-\infty} d\lambda'' n(\lambda'') K(\lambda-\lambda'')F(\lambda'',\lambda'),
\end{eqnarray}
which measures how much the presence of a sea of particles alters the scattering phase between two excitations with rapidities $\lambda$ and $\lambda'$.

The occupation function $n(\lambda)$ defines an energy $\epsilon(\lambda)$ via the relation
\begin{equation}
n(\lambda) = \frac{1}{1+ e^{\epsilon(\lambda)}},\quad \epsilon(\lambda)=\log \frac{\rho_h(\lambda)}{\rho_p(\lambda)} .
\end{equation}
$\epsilon(\lambda)$ can be interpreted as 
a generalized energy contribution measuring the cost of creating an excitation at $\lambda$ around a particular state of the system. 
It can be shown to satisfy the equation
\begin{equation}\label{TBAeq}
\epsilon(\lambda) = \epsilon_0(\lambda) - \int\frac{d\lambda'}{2\pi}K(\lambda-\lambda')\log(1+e^{-\epsilon(\lambda')}).
\end{equation}
$\epsilon_0(\lambda)$, the source term of the above integral equation, can be thought of as the ``bare'' energy of an excitation, what
the excitation energy would be if there were no other excitations in the system.
It is the key quantity for determining how different possible sets of conserved charges describe a particular quench as we discuss in the next section.

\section{Building the GGE with Different Bases of Charges}\label{sec3}

As our starting point for this construction, we suppose that the quench in which we are interested has a known $\epsilon_0(\lambda)$ as defined above. 
Knowing $\epsilon_0(\lambda)$ is equivalent to knowing $\rho_p(\lambda)$ for the quench as we can use 
Eqns. \ref{TBAeq} and \ref{rho_eqn} to go between these two quantities.
$\rho_p(\lambda)$ for a quench can be determined in one of two ways.  It can be determined by using 
the quench action to arrive at a representative state characterized by a given $\rho_p(\lambda)$
or it may be determined by employing the numerical method, NRG+ABACUS, developed to study
quantum quenches \cite{Caux2012,James2017}, to extract the $\rho_p(\lambda)$ associated with a quantum quench.

To see why $\epsilon_0(\lambda)$ is the key quantity for describing the GGE, let us consider the action of the ensemble (\ref{eI}) on a Bethe state:
\begin{equation}
\hat\rho_{GGE}\vert\rho_p\rangle=\frac{1}{Z}e^{-L f[\rho_p]}\vert\rho_p\rangle,
\end{equation}
where $f[\rho_p]$ is the generalized free energy density.  The key point is that $f[\rho_p]$ is given in terms of $\epsilon_0(\lambda)$:
\begin{equation}\label{eps0GGE}
f[\rho_p]=\int d\lambda\epsilon_0(\lambda)\rho_p(\lambda),
\end{equation}
and at the same time is a linear functional of the root density, $\rho_p(\lambda)$. 

A requirement that we will place on our charges, $\{\hat Q_n\}$, is that they involve the root density in the same, linear way,
\begin{equation}\label{charges1}
\hat Q_n\vert\rho_p\rangle=L\int d\lambda q_n(\lambda) \rho_p(\lambda)\vert\rho_p\rangle ,
\end{equation}
where $q_n(\lambda)$ is a function that describes the action of the charge on the Bethe state.
Comparing Eqn. \ref{charges1} with Eqn. \ref{eps0GGE}, we see that
finding a set $\{\hat Q_n\}$ comes down to expanding the coefficient function $\epsilon_0(\lambda)$ on a set of basis functions, $\{q_n(\lambda)\}$, i.e.
\begin{equation}
\epsilon_0(\lambda)=\sum_n \beta_nq_n(\lambda).
\end{equation}
The coefficients of expansion then become the set of generalized inverse temperatures of the GGE.  The so-called ultra-local charges, the charges
that caused difficulties in trying to construct a GGE for the interaction quench in the Lieb-Liniger model \cite{Caux2013,Kormos2013,DeNardis2014}, are given by
\begin{equation}
q_n(\lambda) = \lambda^n.
\end{equation}
Even though the ultra-local charges are not well-defined for the interaction quench, their existence is important for being able to define alternate GGEs.  
As the polynomials provide a complete basis of functions, their existence tells us that we can construct other complete bases of charges (or at least sets
of charges whose associated $q_n(\lambda)$ are locally real analytic in $\lambda$).

From this point of view, finding a set of charges and the associated generalized temperatures is a problem in the domain of approximation theory. All we need to do is to settle on a linear space that includes $\epsilon_0(\lambda)$ and use a complete set of functions in this space to expand it.  If $\epsilon_0(\lambda)$ is not a square
integrable function (as is the case for the interaction quench), i.e.
\begin{equation}
\int^\infty_{-\infty} d\lambda (\epsilon_0(\lambda))^2 = \infty,
\end{equation}
we might want to consider expansion bases that belong to the weighted $L^2$ space, $L^2(\mathbf{R},\omega(\lambda)d\lambda)$, with an appropriate weight function $\omega(\lambda)$.
So, for example, if we suppose our charges to be orthonormal, we would have
\begin{equation}
\int d\lambda\,q_n(\lambda)q_m(\lambda)\omega(\lambda)=\delta_{nm},
\end{equation}
with the corresponding 
generalized temperatures being
\begin{equation}
\beta_n=\int d\lambda\,\epsilon_0(\lambda) q_n(\lambda)\omega(\lambda).
\end{equation}
We now discuss some possible choices of $\{q_n\}$.

We first consider the following set of functions:
\begin{equation}\label{cosexp}
q_0(\lambda)=\frac{1}{2\pi},\qquad q_{n\ge1}(\lambda)=\frac{(-1)^n}{\pi}\cos(2n\arctan\lambda),
\end{equation}
They form an orthonormal set with the weight functions $\omega(\lambda)=\frac{2}{1+\lambda^2}$.
We will see that these functions are well suited to describing a quench characterized by an $\epsilon_0(\lambda)$ with a slight, logarithmic divergence 
as in the interaction quench.  In particular for this quench, the expectations values of the charges on the initial state
are finite, i.e.
\begin{equation}
\int d\lambda q_n(\lambda)\rho_p(\lambda)<\infty,
\end{equation}
and they are even, smooth, and all their derivatives go to zero as $|\lambda|\to\infty$.  As we will see,
this means that they correspond to quasi-local charges, at least for large $c$. 

We also consider using the Chebyshev polynomials:
\begin{align}\label{cheq}
&q_n(\lambda)=c_nT_n(2/\pi\arctan\lambda),\\
&c_0=\frac{1}{\sqrt{\pi}},\quad c_{n>0}=\sqrt{\frac{2}{\pi}},
\end{align}
with the Chebyshev polynomials $T_n(x)$ defined on $-1\le x\le 1$ by
\begin{equation}
T_n(x)=\cos(n\arccos x).
\end{equation}
The associated weight function is $\omega(\lambda)=\frac{1}{(1+\lambda^2)\sqrt{(\pi/2)^2-(\arctan\lambda)^2}}$.
These charges have the same advantages as the ones defined by (\ref{cosexp}).

To demonstrate why we want to consider bases with non-trivial weight functions,  
let us also consider a usual set of orthonormal functions on $-\infty<\lambda<\infty$, the Hermite functions, defined as
\begin{align}\label{Hermch}
q_0(\lambda)&=-(\pi)^{-1/4}e^{-x^2/2},\\
q_n(\lambda)&=(-1)^n(2^n n!\sqrt{\pi})^{-1/2}e^{x^2/2}\frac{d^n}{dx^n}e^{-x^2},\, n>0.
\end{align}
The attendant weight function is $\omega(\lambda)=1$.  We will see that these charges have well-defined expectation values on the initial state for the interaction
quench and that they also correspond to quasi-local operators. 
However, they have exponentially decaying tails and therefore are unable easily to reproduce the $\lambda\to\infty$ behavior of $\epsilon_0$ for the interaction quench.
Since this divergence has to do with the suppression of high energy modes in the representative state, we expect these charges to be suboptimal for this case.

\section{Operatorial Expressions and Quasi-locality of the Charges}

In the previous section we articulated a method for choosing different
sets of conserved charges.  However we do not yet know the operatorial form of these charges.  It is the aim
of this section to provide it.

On the basis of this construction, we will discuss the quasi-locality of the charges.  By quasi-locality we mean that the
charge $\hat Q$ can be expressed as integral over an x-dependent operator $\hat Q_\text{den}(x)$ via
\begin{equation}
\hat Q = \int^\infty_{-\infty} dx\,\hat Q_\text{den}(x),
\end{equation}
where $\hat Q_\text{den}(x)$ is a quasi-local operator, i.e. an operator composed of products of operators, also $x$-dependent, whose
support is primarily confined to the region about $x$.  We will follow Ref. \cite{Fagotti2016} when we allow $\hat Q(x)$ to depend
on operators defined at points far from $x$ provided that this dependence is exponentially small.

The importance of discussing quasi-locality of the charges lays in that it controls, in part, their ability to describe the long time equilibration of the system.
Strictly speaking, when one speaks of equilibration in a closed quantum system, one is concerned about equilibration in
a small part of the system, which we can call A, with the rest of the system, termed B.  The system can be said to come to equilibrium if after we trace out B,
the resulting reduced density matrix, $\hat\rho_A$, equals the GGE density matrix for subsystem A.   However to meaningfully  be able to talk about
the GGE for subsystem A with the same set of charges we associate to the system as a whole, we need the charges forming the GGE to be integrals over 
operators, $\hat Q(x)$, whose support is localized in space. 

To this end, we will show in this section that the Fourier transform of the charge's action on a state, i.e. $q(\lambda)$,
is indicative of how localized its associated charge is. 
In particular we will show that if $\tilde q(x)$ has support that is primarily about 0, we can conclude $\hat Q$ is quasi-local.  As per
Ref. \cite{Fagotti2016},  we will permit the possibility that $\tilde q(x)$ decays exponentially as $x \rightarrow \infty$, and not insist on the more strict condition
that its support around $x=0$ is compact.  The condition that $\tilde q(x)$ is exponentially decaying is ensured by $q(\lambda)$ being even or odd, being smooth, and that all the derivatives go to zero as $\lambda\to\infty$ 
(this can be relaxed to $q^{(n>N)}(\lambda\to\infty)\to0$ with some finite $N$ by allowing distributions), see e.g. \cite{Erdelyi1955}. 

\subsection{$c=\infty$ case}

Let us begin with our demonstration that we can construct charges that are quasi-local in the $c=\infty$ limit.
In this limit the dynamics of the gas become considerably simpler as the interaction kernel $K(\lambda)$ goes to zero,
see for example Eqns. \ref{rho_eqn} and \ref{TBAeq}.  In this limit, the quasi-momenta go to
\begin{equation}
\lambda_j=\frac{2\pi I_j}{L},
\end{equation}
and the Bethe equation reduces to
\begin{equation}
\rho_p(\lambda)+\rho_h(\lambda)=\frac{1}{2\pi}.
\end{equation}
The correspondence between the hard-core bosons and free fermions can be made explicit on the level of operators by a Jordan-Wigner transformation
\begin{equation}\label{string}
\Phi(x)=\exp\left\{-i\pi\int_{0}^x\Psi^\dagger(z)\Psi(z)dz\right\}\Psi(x).
\end{equation}  
Here the hard-core bosonic field $\Phi(x)$ satisfies 
$$
[\Phi(x)^\dagger,\Phi(y)]=0, x\neq y,
$$ 
and 
$$
\Phi(x)\Phi(x)=0,
$$
while the free fermionic field, $\Psi(x)$, satisfies $\{\Psi(x)^\dagger,\Psi(y)\}=\delta(x-y)$.

With these definitions in hand, we now explicitly write $\hat Q$ in terms of the bosonic fields. In terms of the
fermionic fields, $\hat Q$ is given by
\begin{equation}\label{TGcharges}
\hat Q=\sum_\lambda q(\lambda) \Psi^\dagger_\lambda\Psi_\lambda ,
\end{equation}
acting on a Bethe state as
\begin{equation}
\hat Q \vert\lambda_1\ldots\lambda_N\rangle=\sum_{i=1}^N q(\lambda_i)\vert\lambda_1\ldots\lambda_N\rangle,
\end{equation}
that is,
\begin{equation}\label{PsiPsiaction}
\Psi_\lambda^\dagger\Psi_\lambda\vert\lambda_1\ldots\lambda_N\rangle=\sum_{i=1}^N\delta_{\lambda,\lambda_i}\vert\lambda_1\ldots\lambda_N\rangle,
\end{equation}
where in order to avoid unusual normalization factors appearing throughout we choose to remain at finite but large volume $L$. The momentum space operators are defined by
\begin{equation}\label{momspace}
\Psi_\lambda=\frac{1}{\sqrt{L}}\int_0^L dx e^{i\lambda x}\Psi(x),\quad\Psi(x)=\frac{1}{\sqrt{L}}\sum_\lambda e^{-i\lambda x}\Psi_\lambda .
\end{equation}
Using Eqn. \ref{string} we have
\begin{multline}
\hat Q=\int_0^L dx\int_0^L dy \tilde q(x-y) \Phi^\dagger(x)\\ \times\exp\left\{-i\pi\int_y^x\Phi^\dagger(z)\Phi(z)dz\right\}\Phi(y),
\end{multline}
where $\tilde q(x)$ goes to the Fourier transform of $q(\lambda)$ when $L\to\infty$,
\begin{equation}
\tilde q(x)=\int\frac{d\lambda}{2\pi}e^{-i\lambda x}q(\lambda).
\end{equation} 
We can see that the decay in the magnitude of $\tilde q(x)$ away from 0 determines the operatorial spread of the charge density (the integrand above).
As we have discussed above,
if $\tilde q(x)$ is exponentially decaying as $x \rightarrow \infty$, we call $\hat Q$ a quasi-local operator.

For the system of charges defined by $q_m(\lambda)=\cos(2m\arctan(\lambda))$ this exponential decay is present.
It can be inferred from the pole structure of the functions $\cos(2m\arctan\lambda)$: 
they have two m-th order poles at $\lambda=\pm i$. Actually, because of this simplicity, the Fourier transform can be performed analytically, the first few being
\begin{align}
\tilde q_0(x)&=\delta(x);\\
\tilde q_1(x)&=e^{-|x|}-\delta(x);\\
\tilde q_2(x)&=(2|x|-2)e^{-|x|}+\delta(x);\\
\tilde q_3(x)&=(2x^2-6|x|+3)e^{-|x|}-\delta(x);\\
\tilde q_4(x)&=(4/3|x|^3-8x^2+12|x|-4)e^{-|x|}+\delta(x).
\end{align}

For a point of comparison, we consider the ultra-local charges, $q_n(\lambda)=\lambda^n$, and write down their expressions in terms of bosonic fields. We have
\begin{align}
\hat Q_n
&=i^n\int dxdy \delta^{(n)}(x-y) \Psi^\dagger(x)\Psi(y)\\
&=(-i)^n\int dx\Psi^\dagger(x)\partial^{n}_x\Psi(x)\\
&=i^n\int dxdy \delta^{(n)}(x-y) \Phi^\dagger(x)\nonumber\\
&\qquad\times\exp\left\{-i\pi\int_y^x\Phi^\dagger(z)\Phi(z)dz\right\}\Phi(y)\\
&=(-i)^n\int dx\Phi^\dagger(x)\partial^{n}_x\Phi(x).
\end{align}
All the other terms coming from derivatives of the exponential factor disappear because of the hardcore
constraint at $c=\infty$, i.e. $\Phi^2(x)=0$. The remaining term trivially agrees with the results of Refs. \cite{Davies1990,Davies2011}.

\subsection{Operator form of the quasi-local charges at $1/c$}

Having considered the operatorial form of the generalized charges and their quasi-locality at $c=\infty$, we now turn to the case of large but finite $c$.
To construct such a charge we begin by fixing a $q(\lambda)$ that defines a charge $\hat Q$ via
\begin{equation}\label{Qact}
\hat Q\vert\lambda_1\ldots\lambda_N\rangle=\sum_i q(\lambda_i)\vert\lambda_1\ldots\lambda_N\rangle.
\end{equation}
We then suppose that a $1/c$ expansion exists for this $\hat Q$ charge,
\begin{equation}
\hat Q=\hat Q_0+\frac{1}{c}\hat Q_1+O(1/c^2),
\end{equation}
where $\hat Q_0$ is an operator that takes the form
\begin{equation}
\hat Q_0=\sum_\lambda q_0(\lambda) \Psi^\dagger_\lambda\Psi_\lambda,
\end{equation}
i.e. for $c=\infty$, $\hat Q_0=\hat Q$ is conserved and has action
\begin{equation}
\hat Q_0\vert\lambda_1\ldots\lambda_N\rangle=\sum_i q_0(\lambda_i)\vert\lambda_1\ldots\lambda_N\rangle_{c=\infty},
\end{equation}
with $\vert\lambda_1\ldots\lambda_N\rangle_{c=\infty}$ a $c=\infty$ Bethe state.
Our goal then in this section is determine $q_0(\lambda)$ in terms of $q(\lambda)$ and to write $\hat Q_1$
in terms of fermionic operators.

The basic strategy to do this is to insist that $[H,\hat Q]=0$ is satisfied.  To this end 
we employ the fermionic representation of the Lieb-Liniger Hamiltonian.  This has the form
\begin{equation}
H_\text{Lieb-Liniger}=H_0[\Psi]+\frac{2}{c}H_1[\Psi],
\end{equation}
with $H_0$ and $H_1$ \cite{Cheon1999,Yukalov2005,Khodas2007},
\begin{align}
H_0&=-\int_0^L dx \Psi^\dagger(x)\partial_x^2\Psi(x),\\
H_1&=-\int_0^L dx\int_0^Ldy\delta^{(2)}(x-y) \Psi^\dagger(x)\Psi^\dagger(y)\Psi(y)\Psi(x).
\end{align}
(for $H_1$ this expression only holds to order $1/c$ \cite{Brand2005,Cherny2006}).
In terms of the momentum space operators, the Hamiltonian reads 
\begin{align}
H_0&=\sum_\lambda \lambda^2 \Psi^\dagger_\lambda\Psi_\lambda,\\
H_1&=\frac{1}{2L}\sum_{\lambda_1,\lambda_2,\lambda_3} (\lambda_1-\lambda_2)(\lambda_1+\lambda_2-2\lambda_3) \\
&\hskip .5in \times \Psi^\dagger_{\lambda_1}\Psi^\dagger_{\lambda_2}\Psi_{\lambda_3}\Psi_{\lambda_1+\lambda_2-\lambda_3}.
\end{align}
The equality $[H,\hat Q]=0$ then requires
\begin{equation}\label{chargecomm}
[H_0,\hat Q_1]=[\hat Q_0,H_1].
\end{equation}
We immediately see here that $Q_1$ is indeterminate up to an additive $c=\infty$ charge term, i.e. we can equally well redefine 
$\hat Q_1 \rightarrow \hat Q_1 +\delta\hat Q_1$ provided $[\delta \hat Q_1, H_0]=0$. 
For now we will work with a minimal choice $\hat Q_{1\text{min}}$, where no such charge is added and later we will discuss what happens if such a term is added to $\hat Q_{1\text{min}}$. This minimal solution must be in the form of a four-fermion operator like $H_1$,
\begin{equation}
\hat Q_{1\text{min}}=\frac{1}{2L}\sum_{\lambda_1,\lambda_2,\lambda_3} C_{\lambda_1\lambda_2\lambda_3} 
\Psi^\dagger_{\lambda_1}\Psi^\dagger_{\lambda_2}\Psi_{\lambda_3}\Psi_{\lambda_1+\lambda_2-\lambda_3},
\end{equation}
where we used that the total momentum, $\sum_\lambda \lambda\Psi^\dagger_{\lambda}\Psi_{\lambda}$, is conserved at $c<\infty$ as well (up to $1/c^2$ corrections). 
By straightforward calculation we find 
\begin{align}\label{Cformula}
&C_{\lambda_1\lambda_2\lambda_3}=\frac{(\lambda_1-\lambda_2)(\lambda_1+\lambda_2-2\lambda_3) }{\lambda_1^2+\lambda_2^2-\lambda_3^2-(\lambda_1+\lambda_2-\lambda_3)^2}\nonumber\\
&\qquad\hskip -.2in \times\left[q_0(\lambda_1)+q_0(\lambda_2)-q_0(\lambda_3)-q_0(\lambda_1+\lambda_2-\lambda_3)\right],
\end{align}
for distinct $\lambda_1$, $\lambda_2$, $\lambda_3$. For when some or all rapidities coincide, 
$C_{\lambda_1,\lambda_2,\lambda_3}$ is indeterminate since in the commutator (\ref{chargecomm}) the corresponding terms in $Q_1$ are conserved individually in the $c=\infty$ theory,
\begin{equation}
\bigg[\sum_{kl} C_{\kappa\lambda\lambda}\Psi^\dagger_\kappa\Psi^\dagger_\lambda\Psi_\lambda\Psi_\kappa,H_0 \bigg]=0.
\end{equation}

We are now in a position to connect $q_0(\lambda)$ to $q(\lambda)$. This connection will depend on the particular choice of $\hat Q_1$, 
but we will see that the final operatorial  form is independent of this choice. Let us look at the expectation value of the charge using its $1/c$ 
expansion relative to some eigenstate $|\rho_p\rangle$ (whose associated distribution of rapidities is $\rho_p(\lambda )$).  We write this eigenstate in terms of a $1/c$ expansion:
\begin{equation}
|\rho_p\rangle = \vert\rho_{0p}\rangle + \frac{1}{c}\vert1\rangle + \cdots.
\end{equation}
To first order in $1/c$ we then have for the expectation value of $\hat Q$
\begin{widetext}
\begin{equation}\label{chargeevpt}
\langle \rho_p | \hat Q| \rho_p\rangle=\frac{\langle\rho_{0p}\vert \hat Q_0\vert\rho_{0p}\rangle+\frac{1}{c}(\langle 1\vert \hat Q_0\vert\rho_{0p}\rangle+\langle\rho_{0p}\vert \hat Q_0\vert1\rangle+\langle\rho_{0p}\vert \hat Q_1\vert\rho_{0p}\rangle)}{\langle\rho_{0p}\vert\rho_{0p}\rangle+\frac{1}{c}(\langle\rho_{0p}\vert1\rangle+\langle1\vert\rho_{0p}\rangle)}.
\end{equation}
\end{widetext}

The state $|\rho_p\rangle$ can be characterized by assigning to it a set quantum numbers $\{I_j\}$.  For ease we will assume that the total momentum
of $|\rho_p\rangle$ is zero, i.e. $\sum_j I_j = 0$.  These quantum numbers then determine the state's rapidities $\{\lambda_j\}$ via
\begin{equation}
\lambda_j=\frac{2\pi I_j}{L}-\frac{2}{L}\sum_k\arctan\left(\frac{\lambda_j-\lambda_k}{c}\right).
\end{equation}
The state $\rho_{0p}$'s rapidities are then found by taking the $c\rightarrow \infty$ limit of this:
\begin{equation}
\lambda_j=\frac{2\pi I_j}{L}.
\end{equation}
Taking the continuum limit of these equations then leads to a relationship between $\rho_{0p}$ and $\rho_p$:
\begin{equation}
\rho_{0p}(\lambda)=(1-2n/c)\rho_p((1-2n/c)\lambda).
\end{equation}

The first order correction to $|\rho_p\rangle$ can be expressed as a sum of two-particle-hole excitations,
\begin{widetext}
\begin{equation}
\vert1\rangle=\sum_{p_1p_2h_1h_2,\{(p_1,p_2)\neq (h_1,h_2)} D_{p_1,p_2,h_1,h_2}\delta_{p_1+p_2,h_1+h_2} \Psi^\dagger_{p_1}\Psi^\dagger_{p_2}\Psi_{h_1}\Psi_{h_2}
\vert\rho_{0p}\rangle.
\end{equation}
\end{widetext}
Because $\hat Q_0$ is diagonal, the off-diagonal matrix elements $\langle \rho_{0p}|\hat Q_0|1\rangle$ in Eqn. \ref{chargeevpt} vanish.
At this point we are then left with (assuming $|\rho_{0p}\rangle$ has unit normalization):
\begin{equation}\label{chargeevpt2}
\langle \hat Q\rangle=\langle\rho_{0p}\vert \hat Q_0\vert\rho_{0p}\rangle+\frac{1}{c}\langle\rho_{0p}\vert \hat Q_1\vert\rho_{0p}\rangle.
\end{equation}
The minimal $\hat Q_1$ gives the following for the $1/c$ matrix element,
\begin{widetext}
\begin{align}
\langle\rho_{0p}\vert \hat Q_{1\text{min}}\vert\rho_{0p}\rangle
&=\frac{1}{2L}\sum_{\lambda_1,\lambda_2,\lambda_3} C_{\lambda_1\lambda_2\lambda_3}\langle\rho_{0p}\vert\Psi^\dagger_{\lambda_1}\Psi^\dagger_{\lambda_2}\Psi_{\lambda_3}\Psi_{\lambda_1+\lambda_2-\lambda_3}\vert\rho_{0p}\rangle\\
&=\frac{1}{2L}
\sum_{\lambda_1,\lambda_2,\lambda_3} C_{\lambda_1\lambda_2\lambda_3}\langle\rho_{0p}\vert\Psi^\dagger_{\lambda_1}\Psi^\dagger_{\lambda_2}\Psi_{\lambda_3}\Psi_{\lambda_1+\lambda_2-\lambda_3}\vert\rho_{0p}\rangle(\delta_{\lambda_2\lambda_3}-\delta_{\lambda_1\lambda_3})\\
&=\frac{1}{L}\sum_{\lambda_i,\lambda_j}C_{\lambda_i\lambda_j\lambda_j} =L\int d\kappa d\lambda C_{\kappa\lambda\lambda}\rho_{0p}(\kappa)\rho_{0p}(\lambda),
\end{align} 
\end{widetext}
where we have used the antisymmetry of $C_{\lambda_1\lambda_2\lambda_3}$ in its first two arguments. Now the coefficients $C_{\kappa\lambda\lambda}$ 
appearing in the above where the last two rapidities coincide are not fixed in Eqn. \ref{Cformula}.  
If we however require that the charges act on the Bethe states as in Eqn. \ref{Qact}, we can fix this ambiguity. 
Expressing Eqn. \ref{chargeevpt2} in terms of the root densities, $\rho_p(\lambda)$ and $\rho_{0p}(\lambda)$, gives
\begin{align}
\int d\lambda q(\lambda)\rho_p(\lambda)&=\int d\lambda q_0(\lambda)\rho_{0p}(\lambda)\nonumber\\
&\hskip -.1in +\frac{1}{c}\int d\kappa d\lambda C_{\kappa\lambda\lambda}\rho_{0p}(\kappa)\rho_{0p}(\lambda)\\
&=\int d\lambda q_0((1+2n/c)\lambda)\rho_p(\lambda)\nonumber\\
&\hskip -.1in +\frac{1}{c}\int d\kappa d\lambda C_{\kappa\lambda\lambda}\rho_p(\kappa)\rho_p(\lambda)+{\cal O}(1/c^2),
\end{align}
leading to
\begin{equation}
q(\lambda)=q_0((1+2n/c)\lambda)+\frac{1}{c}\int d\kappa C_{\kappa\lambda\lambda}\rho_p(\kappa).
\end{equation}
We however do not want the form of $q(\lambda)$ to depend on the state to which $\hat Q$ is applied. 
This is not allowed by the desired action on Bethe states (\ref{Qact}), therefore the previously arbitrary $C_{\kappa\lambda\lambda}$ has to be chosen to be zero. 

The above argument does not forbid adding a $c=\infty$ charge to $\hat Q_{1\text{min}}$ of the two-fermion form,
\begin{equation}
\hat Q_1=\hat Q_{1\text{min}}+\sum_{\lambda} w(\lambda)\Psi^\dagger_\lambda\Psi_\lambda.
\end{equation}
This modifies the equation for $q_0(\lambda)$,
\begin{equation}
q(\lambda)=q_0((1+2n/c)\lambda)+\frac{1}{c}w(\lambda),
\end{equation}
which upon inversion gives
\begin{equation}
q_0(\lambda)=q((1-2n/c)\lambda)-\frac{1}{c}w(\lambda).
\end{equation}
But this means that the $w$-charge added to $\hat Q_{1\text{min}}$ will cancel out from $\hat Q$ because the same term with the opposite sign has to be added to $\hat Q_0$ 
as well. Therefore, we arrive at the unique expression for the charge $\hat Q$:
\begin{widetext}
\begin{equation}\label{finQ}
\hat Q=\sum_\lambda q(\lambda(1-\frac{2n}{c}))\Psi^\dagger_\lambda\Psi_\lambda+\frac{1}{2Lc}\sum_{\lambda_1,\lambda_2,\lambda_3} C_{\lambda_1\lambda_2\lambda_3}\Psi^\dagger_{\lambda_1}\Psi^\dagger_{\lambda_2}\Psi_{\lambda_3}\Psi_{\lambda_1+\lambda_2-\lambda_3}+{\cal O}(1/c^2),
\end{equation}
with
\begin{equation}
C_{\lambda_1\lambda_2\lambda_3}=\frac{(\lambda_1-\lambda_2)(\lambda_1+\lambda_2-2\lambda_3)\left[q(\lambda_1)+q(\lambda_2)-q(\lambda_3)-q(\lambda_1+\lambda_2-\lambda_3)\right]}{\lambda_1^2+\lambda_2^2-\lambda_3^2-(\lambda_1+\lambda_2-\lambda_3)^2},\quad C_{\lambda_1\lambda_2\lambda_2}=0.
\end{equation}

\subsubsection{Locality of $1/c$ terms}

We now turn to the locality of the charge we have constructed in a $1/c$ expansion.
Rewriting the expression for the $1/c$ corrections of the charge in
terms of real space operators (\ref{momspace}) we arrive at (in the
$L\to\infty$ limit -- see Appendix A):
\begin{eqnarray}\label{realspace}
\hat Q &=& \int_0^Ldx\int_0^Ldy\left((1+\frac{2n}{c})\tilde q((1+\frac{2n}{c})(x-y))\right)\Psi^\dagger(x)\Psi(y)\cr
&& + \frac{1}{2c}\int_0^Ldx_1\int_0^Ldx_2\int_0^Ldx_3\int_0^Ldx_4 F(x_1,x_2,x_3,x_4) \Psi^\dagger(x_1)\Psi^\dagger(x_2)\Psi(x_3)\Psi(x_4);\\
F(x_1,x_2,x_3,x_4) &=& \int\frac{d\lambda_1}{2\pi}\int\frac{d\lambda_2}{2\pi}\int\frac{d\lambda_3}{2\pi}C_{\lambda_1\lambda_2\lambda_3}e^{i\lambda_1x_{41}}e^{i\lambda_2x_{42}}e^{i\lambda_3x_{34}}\cr\cr
&&\hskip -.85in = \frac{1}{4}\tilde q(x_3+x_4-x_1-x_2)\bigg[(\text{sgn}x_{23}-\text{sgn}x_{41})(\delta'(x_{42})-\delta'(x_{13}))+(\text{sgn}x_{42}-\text{sgn}x_{13})(\delta'(x_{41})-\delta'(x_{23}))\bigg],
\end{eqnarray}
with $x_{ij}=x_i-x_j$. Using this expression it is easy to check that the integrand of $F(x_1,x_2,x_3,x_4)$ becomes exponentially small when any of the $x_i$'s diverges from
any of the other $x_i$'s.  And while we have expressed the charges at $1/c$ in terms of the fermions, they are similarly quasi-local in the bosonic description
as the string operators are confined to run between the $x_i$.

\end{widetext}

\section{Interaction quench in the Lieb-Liniger model}

Now we will apply the ideas developed in the previous section to the interaction quench in the LL model (\ref{LLHam}). This protocol 
refers to taking the ground state of (\ref{LLHam}) at interaction strength $c_0=0$ and studying the dynamics under (\ref{LLHam}) at
some finite repulsive interaction strength $c$. 

For this quench 
an exact formula is available describing $\epsilon_0(\lambda)$ \cite{DeNardis2014}, the key quantity for our purposes as discussed in Section \ref{sec3}:
\begin{equation}\label{qavarep}
\epsilon_0(\lambda)=2\log(c/n)+\log\left[\left(\frac{\lambda}{c}\right)^2\left(\left(\frac{\lambda}{c}\right)^2+\frac{1}{4}\right)\right].
\end{equation}
This coefficient function diverges only logarithmically in $\lambda$, which in turn corresponds to the density of particles $\rho_p(\lambda)$ having a polynomial tail in $\lambda$, i.e. 
$\rho_p(\lambda)\sim \lambda^{-4}$, $\lambda\to\infty$ \cite{DeNardis2014}, making the ultra-local charges ill-defined on this state for $n>3$, i.e. $E^{n>3}=\infty$
\cite{Kormos2013}.

\begin{figure}[ht]
\includegraphics[scale=.9]{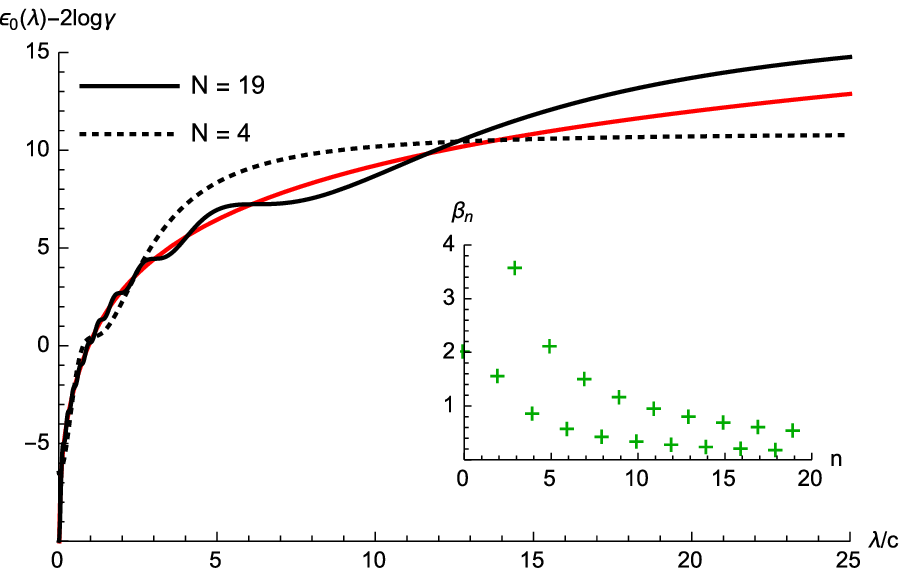}
\includegraphics[scale=.9]{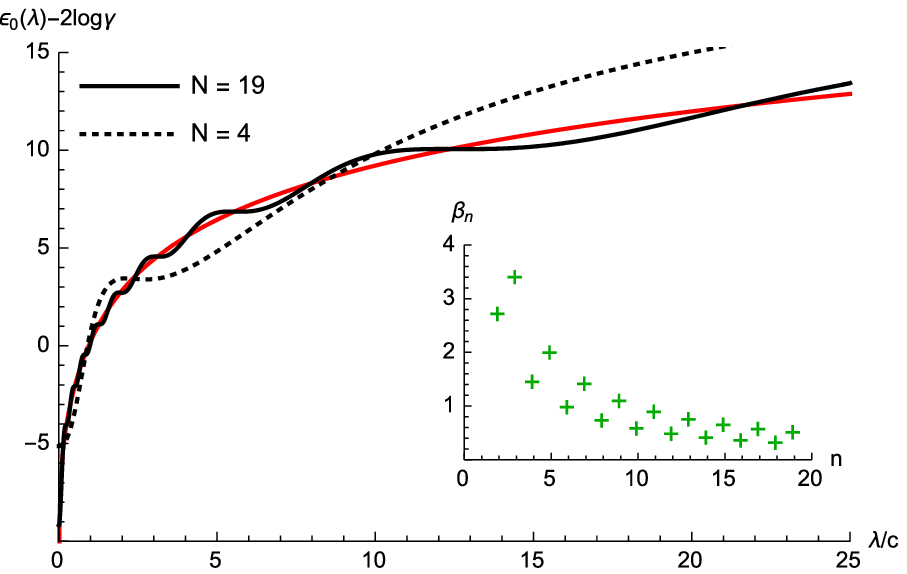}
\includegraphics[scale=.9]{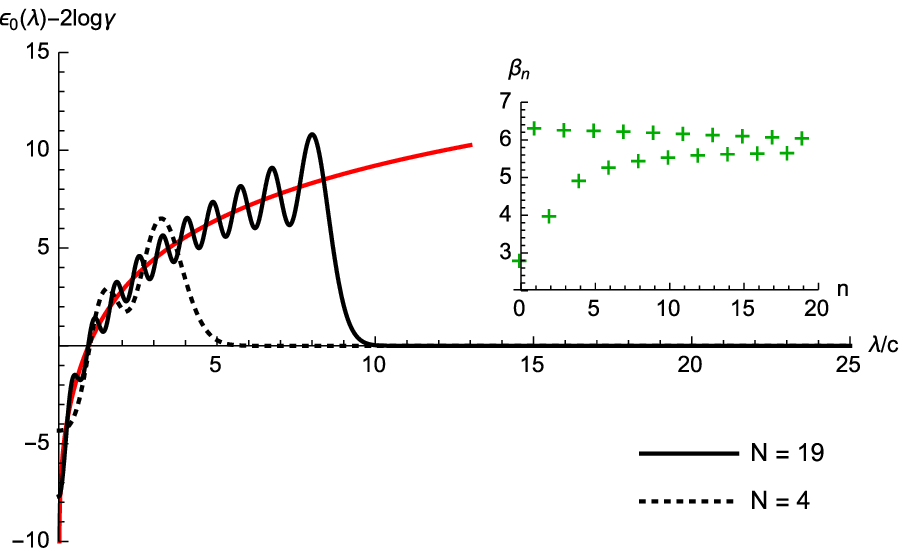}
\caption{Approximations of $\epsilon_0(\lambda)$ (red line)
using the transformed cosine (\ref{cosexp}) (top panel), Chebyshev (\ref{cheq}) (middle panel) and Hermite (\ref{Hermch}) (bottom panel) charges
where we have truncated the GGEs to 5 and 20 charges and assuming a finite $c$. 
For the Chebyshev polynomials the vanishing odd charges are not counted.
In the insets of the three panels, the attendant generalized temperatures are plotted.}\label{fig1}
\end{figure}

Unlike the ultra-local charges, the three sets of charges defined in Section 3 (Eqns. \ref{cosexp}, \ref{cheq}, and \ref{Hermch})
have finite expectations on the initial state of the interaction
quench and correspondingly provide a good basis for expanding $\epsilon_0(\lambda)$.
In Fig. \ref{fig1} we show expansions of $\epsilon_0(\lambda)$ truncated to a finite number of charges, $\epsilon_0(\lambda)=\sum_{i=0}^N\beta_iq_i(\lambda)$, using these three families of charges (\ref{cosexp}), (\ref{cheq}) and (\ref{Hermch}). We also show 
the corresponding generalized temperatures (the coefficients of expansion) in the insets of this figure.
For the transformed cosine and Chebyshev charges, the expansion converges rapidly.  Including only 5 charges in the expansion already provides a decent
approximation to $\epsilon_0(\lambda)$.  We also see for these two cases the generalized inverse temperatures decay rapidly in size with increasing
charge index.  In contrast the expansion of $\epsilon_0(\lambda)$ with the 
Hermite charges is not uniform for all $\lambda$.  We also see that
the Hermite generalized temperatures are not obviously tending towards
zero.  This is an indication that 
$\epsilon_0(\lambda)$ is not square integrable with the weight $\omega(\lambda)=1$. 
Ultimately however, the true measure of a truncated GGE based upon a particular set of charges is the quality of reproduction of physical quantities, i.e. some parts of $\epsilon_0$ 
will be more important for the physics than others. This will be discussed in the next section.

\subsubsection{Alternate Determination of Generalized Temperatures}

In Sec. 3 we described a straightforward method of finding the generalized temperatures in the GGE once a system of charges is defined: 
we expanded the source term, $\epsilon_0(\lambda)$, of the generalized free energy on the  functions describing the charges in some well defined space of 
square integrable functions. This process requires knowledge of $\epsilon_0(\lambda)$, which may not always be available.
In this subsection we will therefore consider an alternative method of finding the generalized temperatures: comparing the expectation values of the charges 
in the initial state and in a truncated GGE.

In this alternate procedure to determine the generalized temperatures, we suppose that we are given as input the expectation
values post-quench of the conserved charges.  
To then find the generalized temperatures $\beta_i$ in $\epsilon_0(\lambda)=\sum_i\beta_iq_i(\lambda)$, we will solve the following set of nonlinear equations:
\begin{equation}\label{charges}
\langle \hat Q_i\rangle=\int \frac{d\lambda}{2\pi}q_i(\lambda)\frac{1}{1+e^{\sum_{n=0}^N\beta_nq_n(\lambda)}},\quad i=0,\ldots,N .
\end{equation}
Note that solving such a system of nonlinear equations, 
especially for a large number of generalized temperatures, can be challenging. In fact, we found that the solution is in general not unique and 
to get the right one we had to use some information available through expanding $\epsilon_0(\lambda)$ on $q_i(\lambda)$ to set the initial values 
of the iterative solution scheme. An alternative, more stable method based on exploiting fluctuation-dissipation
relations to obtain the generalized temperatures was proposed in Ref. \cite{Foini2017,deNardis2017}.
Assuming for now that we can find the right solution for (\ref{charges}), then to get $\rho_p(\lambda)$ we solve Eqs. (\ref{TBAeq}) and (\ref{rho_eqn}) consecutively.

\begin{figure}[h]
\includegraphics[scale=0.9]{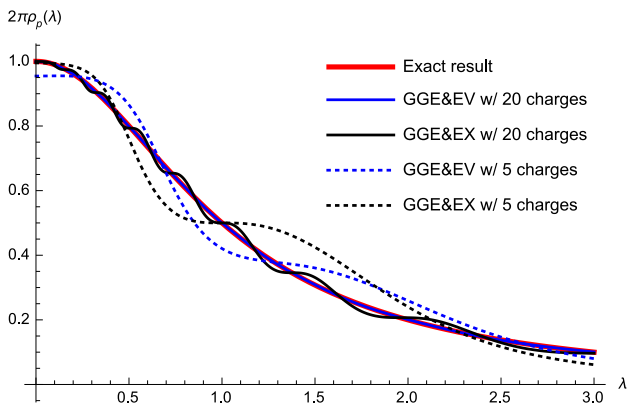}\caption{Here are given reconstructions of the mode occupation number density in the $c=\infty$ limit using two
different means to determine the generalized temperatures.  In the first, labeled GGE\&EV, we find the temperatures by performing a fit using Eqn. \ref{charges}
to the known expectation values of the charges.  In the second, labeled GGE\&EX, we read off the temperatures by expanding $\epsilon_0(\lambda)$
in the basis of charges.  Here we compare both methods for two different truncated GGEs: one where we keep the first 5 charges and one where we keep 20.
We work with a density of the gas of $n=1/2$ and so the exact rapidity density is given by $\rho_p(\lambda) = \frac{1}{2\pi}\frac{1}{1+\lambda^2}$ and only $\hat Q_0$ has a non-vanishing expectation value.}
\label{figreco}
\end{figure}

In Fig. \ref{figreco} we compare reconstructions of the mode occupation density $2\pi\rho_p(\lambda)$ in the BEC-to-TG protocol obtained from the
two different methods to determine the generalized temperatures for the transformed cosine charges.  These two methods are i) 
truncated expansions of $\epsilon_0(\lambda)$ (denoted by 'GGE\&EX') and 
ii) fitting the parameters of the GGE to the expectation values of charges via Eqn. \ref{charges} (termed GGE\&EV).  
We see that when we perform the reconstruction with a small number (5) of charges, the two reconstructions agree (roughly) equally well
with the exact form of $\rho_p(\lambda)$.  However when we expand the number of charges to 20, we see that the GGE\&EV method for determining
the temperatures leads to almost perfect agreement between $\rho_p(\lambda)$ and its reconstruction.  However for the GGE\&EX method, 20 charges
still leads to noticeable deviations.

\begin{figure}[h]
\includegraphics[scale=0.9]{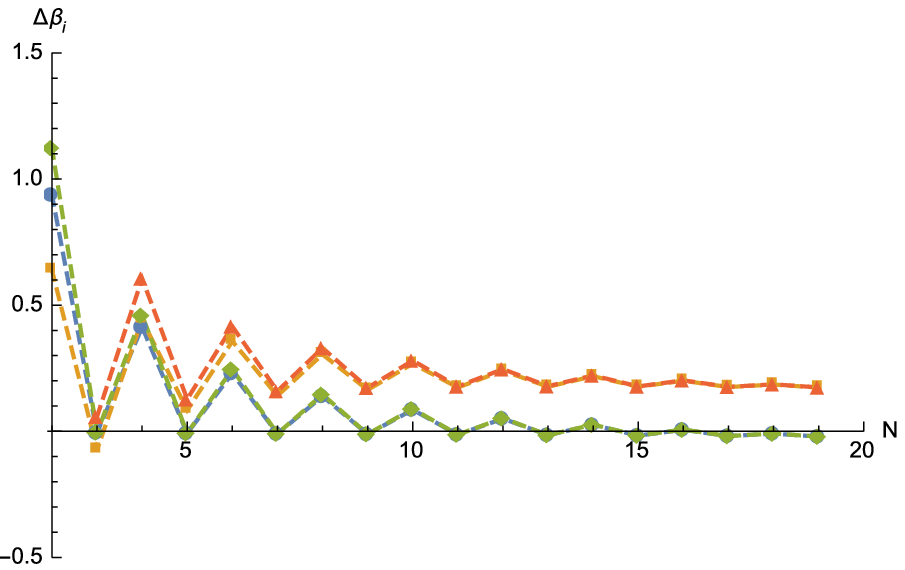}\\
\includegraphics[scale=0.9]{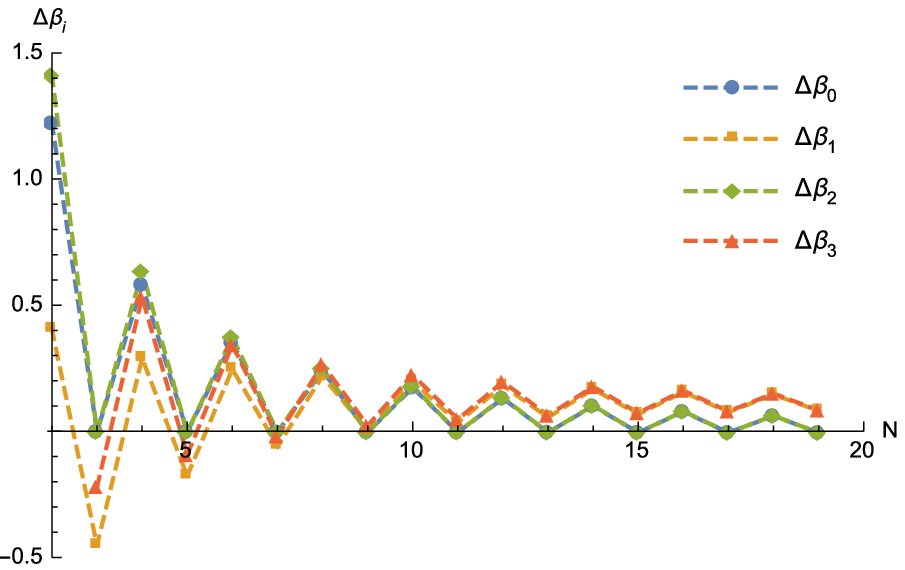}
\caption{Differences between the GGE\&EV values of the first four generalized 
temperatures for the transformed cosine charges and their GGE\&EX values as a function of the truncation $N$. The cutoff in the integral (\ref{charges}) was set $\Lambda=50$ (upper panel) and $\Lambda=100$ (lower panel).}
\label{betafig}
\end{figure}

An important question here is how the temperatures as determined in the GGE\&EV method converge to their GGE\&EX counterparts as the number
of charges in the (truncated) GGE is increased (or whether they converge at all).  
In Fig. \ref{betafig} we show the dependence of the first four generalized temperatures on the truncation $N$ obtained in the GGE\&EV scheme relative to their GGE\&EX 
values: $\beta_0=\beta_2=0$, $\beta_1=4$, $\beta_3=4/3$. The two schemes to determine the generalized temperatures give different reconstructions using the 
same number of charges, however in the $N\to\infty$ limit the GGE\&EV should converge to the GGE\&EX values. We however see from Fig. \ref{betafig} that after
a certain $N$ the approach of the two values cease.  This happens because when we are 
solving the nonlinear equations, we have truncated the integral to a finite domain $-50<\lambda<50$ and $-100<\lambda<100$, respectively. We verified that increasing this cutoff starts to slowly 
decrease the $N\to\infty$ differences between the two methods.

\subsubsection{Density-density correlation function from the truncated GGE}

As we have indicated, an important measure of how efficient a truncated GGE is its efficacy in describing 
physical quantities in the post-quench system.  To this end we consider the density-density correlation function,
both its time independent and time dependent variants.

\begin{figure}[h]
\includegraphics[scale=0.9]{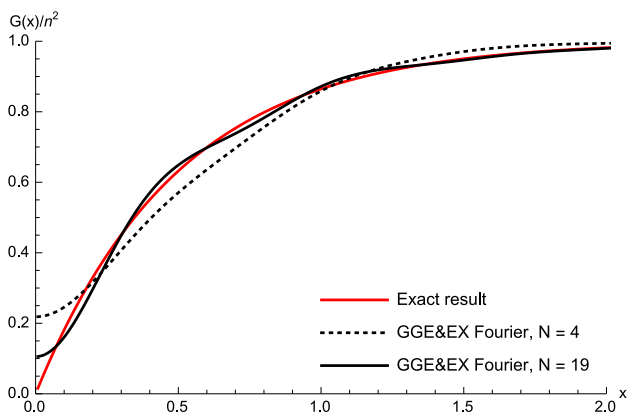}\caption{Density-density correlation function in the TG limit at $n=1/2$ from the truncated GGE using 5 and 20 transformed cosine charges.}\label{fig3}
\end{figure}

\begin{figure}[h]
\includegraphics[scale=0.9]{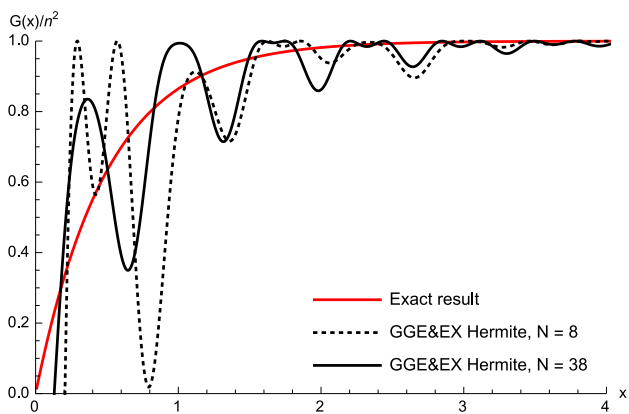}\caption{Density-density correlation function in the TG limit at $n=1/2$ from the truncated GGE using 5 and 20 transformed Hermite function charges. (The odd charges with vanishing temperatures are not counted.)}\label{fig4}
\end{figure}

We begin by looking at the time independent case in the TG limit:
\begin{multline}
G(x)=\langle\rho_p\vert\Phi^\dagger(x)\Phi(x)\Phi^\dagger(0)\Phi(0)\vert\rho_p\rangle\\=n^2-\left(\int d\lambda e^{ix\lambda}\rho_p(\lambda)\right)^2,
\end{multline}
and compare its reconstructions using different truncations of the charges (\ref{cosexp}). The above formula can easily be proved 
using (\ref{PsiPsiaction}) \cite{DeNardis2014,Imambekov2009}. (The density-density correlation function can also be obtained in the low energy limit, see \cite{Eriksson2013}).
In Fig. \ref{fig3} we show results for the reconstruction of $G(x)$ using 5 and 20 charges whose temperatures
are determined in the GGE\&EX scheme. In addition to the well-behaving transformed cosine charges (\ref{cosexp}),
we also display results using the Hermite function charges (\ref{Hermch}) in Fig. \ref{fig4}. 
Reconstructions in the latter case are far inferior to the former one, as expected.

\begin{figure}[ht]
\includegraphics[scale=0.9]{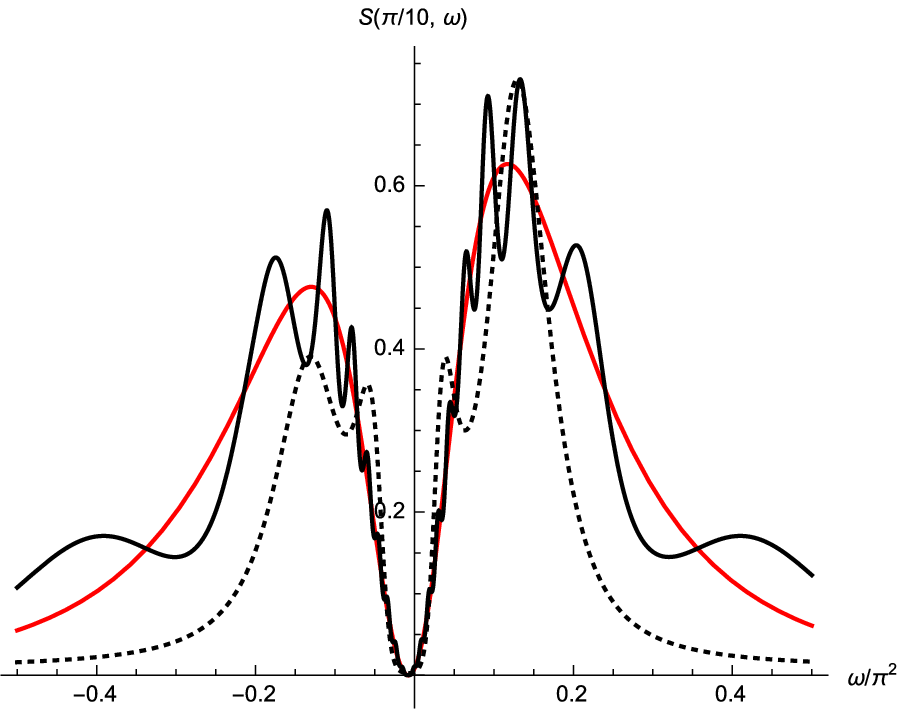}
\\
\includegraphics[scale=0.9]{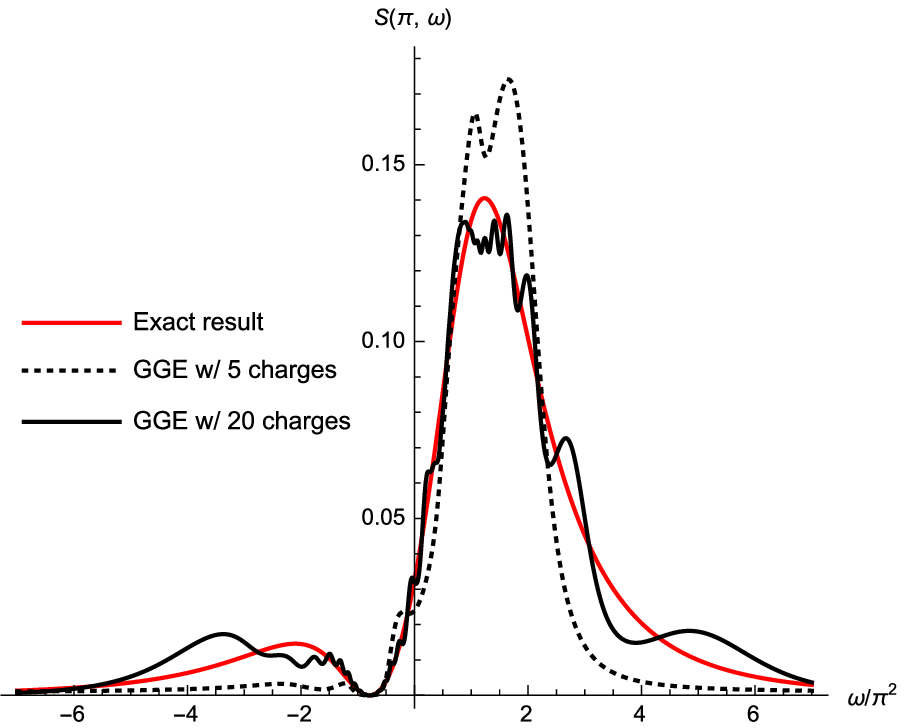}
\caption{DSF to first order in $1/c$ at $q=\pi/10$ (top panel) and $q=\pi$ (bottom panel) 
for the representative state after the interaction quench $c=0$ to $c=16$ (red lines) and their approximations using 5 (dotted lines) and 
20 charges (black lines) of the transformed cosine charges, $q_m(\lambda)=\cos(2m\arctan(\lambda))$.}\label{figSqo}
\end{figure}

We now turn to the time dependent density-density correlation function or dynamic structure factor (DSF),
\begin{equation}
S(q,\omega)=\int dx dt e^{iqx-i\omega t}\langle\rho_p\vert\hat\rho(x,t)\hat\rho(0,0)\vert\rho_p\rangle,
\end{equation}
as obtained from different reconstructions of the representative state. A formula is available for the DSF in the $c\gg1$ limit (taking here $n=1$)
\cite{DeNardis2015},
\begin{multline}
S(q,\omega)=\left(\frac{1+6/c}{2q}+\frac{1}{\pi c}\fint d\lambda\frac{n(\lambda+p)-n(\lambda+h)}{\lambda}\right)\\ \times n(h)(1-n(p)),
\end{multline}
where $n(\lambda)$ is the filling function, $n(\lambda)=\rho_p(\lambda)/(\rho_p(\lambda)+\rho_h(\lambda))$ and the rapidities $\lambda=p$ and $\lambda=h$ 
describe the corresponding particle-hole excitation, $q=(1+2/c)(p-h)$ and $\omega=p^2-h^2$ to first order in $1/c$ or
\begin{align}
p=&\frac{q}{2(1+2/c)}+\frac{\omega(1+2/c)}{2q};\\
h=&-\frac{q}{2(1+2/c)}+\frac{\omega(1+2/c)}{2q}.
\end{align}
To exploit the DSF formula at $c$ large,
we need to determine the filling function $n(\lambda)$ that corresponds to a truncated GGE.  This can be done  numerically by expanding
$\epsilon_0(\lambda)$ in terms of the charges and then solving (\ref{TBAeq}) for $\epsilon=\log(\rho_h/\rho_p)$ -- here $\epsilon_0(\lambda)$ serves
as a source term.  Solving this equation is done easily by iteration in Fourier space.  The principal value integral in the above expression for $S(q,\omega)$
can easily been evaluated after subtracting the pole contribution at $\lambda=0$, which in any case we found to be heavily suppressed for small $q$. 
Results of these calculations are shown in Fig. \ref{figSqo}.

\begin{figure*}[ht]
\includegraphics[scale=0.9]{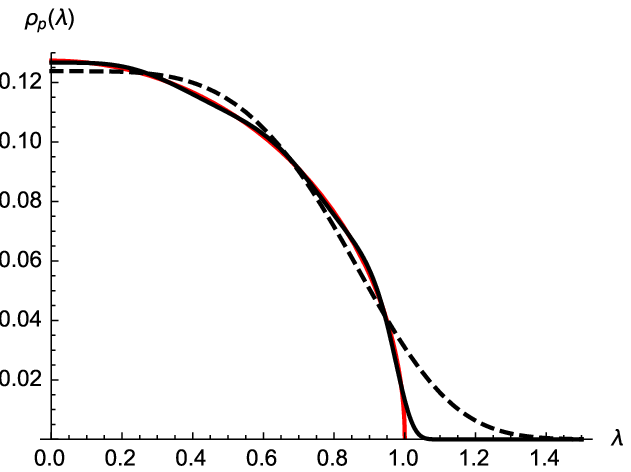}
\includegraphics[scale=0.9]{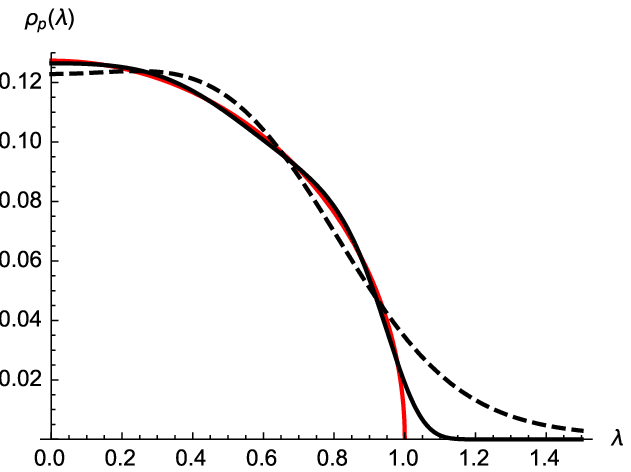}
\includegraphics[scale=0.9]{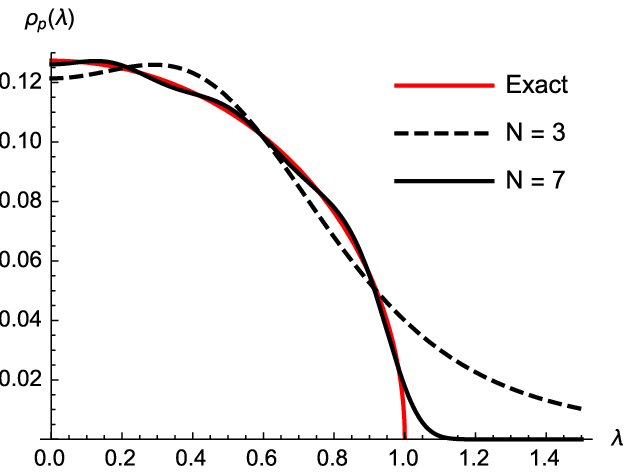}
\caption{Reconstructions of $\rho_p(\lambda)$ for the trap release quench using the ultra-local (left panel), quasi-local Hermite (middle panel) and cosine (right panel) charges 
using the GGE\&EX method at truncations $N=3$ and 7. The parameters of this quench were taken as $n=0.2$ and $\nu=1$.}\label{figtrap}
\end{figure*}

\section{Trap release}

While the ultra-local charges are ill-defined for the interaction quench, there are, of course quench protocols where they can be sensibly used.  It is interesting
in such cases to compare the ultra-local charges with the quasi-local ones to see which perform better.  (This is a question that animated Ref. \cite{Essler2017}.) We will consider this
question in the context of the release of the Lieb-Liniger gas from a harmonic trap. The initial state is the ground state of the Hamiltonian  
\begin{equation}
H=-\sum_{j=1}^N\frac{\partial^2}{\partial x_j^2}+\frac{1}{4}\sum_{j=1}^N\omega^2x_j^2+2c\sum_{i\neq j}\delta(x_i-x_j)
\end{equation}
and the dynamics is governed by the Lieb-Liniger Hamiltonian (\ref{LLHam}), i.e. the above with $\omega =0$.
For this quench we will compare the performance of truncated GGEs based on the ultra-local and quasi-local charges.

In \cite{Collura2013,Collura2013a} the equilibration of the Tonk-Girardeau gas released from a trap was studied.  There the gas was studied 
in the thermodynamic limit, i.e. $N,L \rightarrow \infty$, with
particle density $n=N/L$ fixed, but with the condition that $\nu =\omega N$ was kept constant.
They found that $\rho_p(\lambda)$ for this quench is given by
\begin{equation}\label{nktrap}
\rho_p(\lambda)=\frac{2n}{\pi\nu}\sqrt{1-(\lambda/\nu)^2}
\end{equation}
The true tail of $\rho_p(\lambda)$ is a Gaussian instead of the sharp cutoff at $\lambda = \nu$ for finite but small $n$. 
It is clear that this expression only makes sense for $\nu>2^{3/2}n$ (otherwise the filling $2\pi\rho_p(\lambda)$ would be greater than 1 near $\lambda=0$).
This condition amounts to insisting the size of the gas in its trapped initial state is smaller than the size of the system so that there is an actual
expansion of the gas once released.

In order to test which set of charges form a better truncated GGE using the same number of charges, we used the GGE\&EX method and solved Eqs. (\ref{charges}) for the transformed cosine, the Hermite, and the ultra-local charges 
at different truncations. The recovery of the root density is shown in Fig. \ref{figtrap}. 
In this case it is the truncated sets of ultra-local charges that best reproduce the exact $\rho_p$.

\section{Discussion}

We have presented herein a discussion of how one can construct
arbitrary bases of conserved charges in integrable models.   These bases can be tailored so as to allow them to describe
in an efficient manner particular quantum quenches (in the sense that one can write down a GGE density matrix for the post-quench state of the system).  One example that we have
focused on in this paper is the interaction quench in the Lieb-Liniger model.  As we have already discussed, in this quench the standard ultra-local charges fail to describe the 
quench \cite{Kormos2013,DeNardis2014}.  In this quench, excitations
are created at arbitrarily high momenta and so only the first three of
the ultra-local charges have a finite value after the quench.  We however have shown how to construct quasi-local charges that have finite expectation values 
for this particular quench.

In constructing these quasi-local charges, we do not work directly with operatorial expressions for the charges.  Rather we work with the quantity $\epsilon_0(\lambda)$, the source
term of the pseudo-energy equation in Eqn. \ref{TBAeq} (and for the interaction quench given explicitly in Eqn. 25).  For our purposes this quantity is primary as it describes
the action of the GGE density matrix on a Bethe state $|\{\lambda_i\}\rangle$ (where the $\lambda_i$'s are solutions to the Bethe equations) via
\begin{equation}
\hat\rho_\text{GGE}|\{\lambda_i\}\rangle = e^{-\sum_i \epsilon_0(\lambda_i)}|\{\lambda_i\}\rangle.
\end{equation}
Thus by expanding the function $\epsilon_0 (\lambda)$ in terms of a complete set of functions $\{q_i(\lambda ) \}$, i.e.
\begin{equation}
\epsilon_0(\lambda) = \sum_i \beta_i q_i(\lambda),
\end{equation}
one can arrive at different sets of charges where the $\beta_i$'s are the different inverse temperatures and the $q_i(\lambda)$ describe the
action of the charges $\hat Q_i$ on the Bethe states:
\begin{equation}
\hat Q_j |\{\lambda_i\}\rangle = (\sum_iq_j(\lambda_i))|\{\lambda_i\}\rangle.
\end{equation}
And as we showed in Sections IV, the quasi-locality of the charges is directly correlated with the support of the Fourier transform of
$q_i (\lambda)$.

The locality property of the sum of all the operators defined in this way, i.e. that of the log of the GGE operator, is controlled however by the locality of $\epsilon_0$'s Fourier transform. 
Equivalently we could inquire about the locality of the charge defined by $\epsilon_0$ itself. 
In case of the interaction quench, the Fourier transform of such a charge has a $1/|x|$ tail, signaling non locality. So while the individual charges that we utilize are always quasi-local, we are actually trying to approximate a non-local operator here. This has been discussed in the context of the different quench, for the XXZ spin-chain model already, in \cite{Ilievski2017,Pozsgay2017}. We note that the non-locality of $\epsilon_0$ might have implications for the thermalization of local observables, as we expect that local observables might only thermalize via local GGEs.

One practical advantage of our construction of quasi-local charges over the original ultra-local charges 
is that we can employ bases of charges, $\{\hat Q_i \}$ whose action on the Bethe states $\{q_i(\lambda)\}$ is bounded in value as the value of $\lambda \rightarrow \infty$.
While of course this is necessary if one is to construct a GGE for the interaction quench in Lieb-Liniger, it makes one's life numerically easier
in studying arbitrary quenches.  As one example, in Ref. \cite{Brandino2015} a quench of a 1D Bose gas prepared in a parabolic potential and then released
into a cosine potential was considered.  The aim here was to demonstrate that even though the post-quench Hamiltonian broke integrability, a remnant
of the conserved charges survived (at finite particle number).  Doing so however was made more difficult by the use of the ultra-local charges.  Because the construction
used ultra-local charges $Q_{n}$ whose action on a Bethe state was
\begin{equation}
Q_n |\{\lambda_i\}\rangle = \sum_i \lambda^n_i |\{\lambda_i\}\rangle,
\end{equation}
one had to deal with charges that took large numerical values.  This construction would have been easier if a quasi-local
set of operators whose action on the Bethe states was finite had been available at the time.

This work extends the notion of quasi-local charges discussed in Refs. \cite{Doyon2004,Essler2015,Essler2017} in the context of the free fermionic field
theoretic representation of the quantum Ising model.  The discussion here took a different tack than taken there.  In \cite{Doyon2004,Essler2015,Essler2017},
the operatorial expressions of the charges, $I(\alpha)$, were written down first and the corresponding action of the charges then determined.  These charges were parameterized
by a single positive real variable $\alpha$ controlling their locality (the range of the associated charge density operator equals $\alpha$).  Equivalents
to these charges do exist in our case for $c=\infty$, the analog being
\begin{eqnarray}
\hat Q_{\cos \alpha}  |\{\lambda_i\}\rangle &=& \sum_i \lambda_i^2\cos(\alpha\lambda_i);\cr\cr
\hat Q_{\sin \alpha}  |\{\lambda_i\}\rangle &=& \sum_i \sin(\alpha\lambda_i).
\end{eqnarray}
This is perhaps the most natural basis of expansion of $\epsilon_0(\lambda)$, that of a Fourier integral.  And these charges have finite
expectation values for the interaction quench.  Perhaps their only drawback is that this basis is not discrete ($\alpha$ is a continuous
variable) and one thus needs a strategy to choose a finite number of them in implementing a truncated GGE (but see \cite{Essler2017} for such a procedure).

Our approach to forming different GGEs includes the particular GGE presented in Ref. \cite{Ilievski2017}.  In this work the authors advocate forming a GGE density matrix
which takes the form (in the context of the Lieb-Liniger model),
\begin{equation}
\hat\rho_\text{GGE} |\{\lambda_i\}\rangle= \frac{1}{\cal Z} \exp\bigg[ {\int d\lambda \epsilon_0(\lambda) \hat\rho_p(\lambda)}\bigg]|\{\lambda_i\}\rangle,
\end{equation}
exactly the starting point of this paper.
Having written $\hat\rho_\text{GGE}$ in this form, the differences between Ref. \cite{Ilievski2017} and our work begin to appear however.  
The authors consider their conserved charges in the theory as coming from 
the operator $\hat\rho_p(\lambda)$ which acts, in the thermodynamic limit, on a Bethe state, $|\{\lambda_i\}\rangle$,
via
\begin{equation}
\hat\rho_p(\lambda) |\{\lambda_i\}\rangle = \rho_p(\lambda) |\{\lambda_i\}\rangle,
\end{equation}
i.e. this operator has as its eigenvalues the density of excitations at $\lambda$.  This differs from our approach in two ways.  We instead treat
$\epsilon_0 \equiv \hat\epsilon_0 $ as an operator, or more precisely a linear combination of quasi-local operators whose coefficients of expansion
are the generalized inverse temperatures.  The underlying motivation is also different.  For Ref. \cite{Ilievski2017}, the introduction of $\hat\rho_p(\lambda)$ as
a continuum set of conserved charges is done in the context of a specific model, the XXZ Heisenberg spin chain.  There it is known that one needs, in general, to employ not just
the ultra-local charges, but an infinite set of families of charges $\{X_s(\lambda)\}$, that can be found from a set of generalized transfer matrices built using higher 
spins, $s$, in the framework of the algebraic Bethe Ansatz.  In Ref. \cite{Ilievski2017} it was shown that it was not possible generically to write down a GGE in terms of these
charges and so they proposed as an alternative the family of charges, $\{\hat\rho_p (\lambda)\}$, which while non-local (at least for the case of 
the Lieb-Liniger -- for the XXZ Heisenberg spin chain see the discussion in \cite{Ilievski2017,Pozsgay2017}), do enable one
to write down a GGE for the XXZ Heisenberg model.  (We do note parenthetically that if one is willing to represent the GGE as the limit of a set of truncated GGEs,
the technical difficulty identified by Ref.\cite{Ilievski2017} is avoided, a fact
established in Ref. \cite{Pozsgay2017}.)
Our motivation is however different.  We are interested in finding bases of quasi-local charges that are optimized for different quenches. This is where the practical aspects of our work differs from what was done in Ref. \cite{Pozsgay2017}, where $\epsilon_0(\lambda)$ was expanded on specific orthogonal linear combinations of a truncated set of specific charges, including the ultra-local ones.

Despite these differences, the finding of Ref. \cite{Ilievski2017} is interesting -- namely that there exists complete bases of conserved charges where it is not possible
to write down a density matrix involving those charges for an arbitrary quantum quench.  It is thus worthwhile asking whether this is the case for Lieb-Liniger model.
Here the answer would seem to be no. The problem identified by Ref. \cite{Ilievski2017} 
could then most likely be associated with a more complicated particle content as the Lieb-Liniger model admits a single particle species.
Where such difficulties might show up is any model with string solutions to the Bethe equations (e.g. \cite{Piroli2016,Vernier2017,Mestyan2017}), including quenches that involve multi-component Lieb-Liniger systems such as \cite{PhysRevLett.116.145302,Mathy2012,PhysRevA.89.063627,PhysRevA.89.041601}.

As we have discussed the findings of Ref.  \cite{Ilievski2017}, it is worthwhile also to consider a related construction of a set of conserved charges.  In the $c=\infty$ limit, an oft used set of charges are associated with the occupation numbers \cite{Rigol2007}. The occupation number charges, $\hat n_I$, have expectation values between 0 and 1 and mark when there is a particle with momentum, 
$$
\lambda = \frac{2\pi I}{L},
$$
where the quantum number $I$ is a half-integer/integer (see Eqn. \ref{bethe}).  Using $\epsilon_0(\lambda)$ and Eqn. \ref{bethe1} we can generalize this notion 
away from $c=\infty$.  At $c=\infty$ there is a simple relationship between the momenta, $\lambda_I$, and the quantum numbers $I$.   While at finite $c$,
this relationship becomes more complex, it is still possible to write it down as we have done in Eqns. \ref{bethe1} and \ref{F}.  If $\lambda(I)$ is the momentum determined
by the quantum number $I$ as determined by Eqn. \ref{bethe1}, the expectation value of the occupation number operator is
\begin{equation}
\langle \hat n_I \rangle = \frac{1}{1+e^{\epsilon (\lambda (I))}} .
\end{equation}
We can easily write the GGE associated with these charges by writing the action of the density matrix on a Bethe state
\begin{eqnarray}
\hat\rho_\text{GGE} |\{\lambda(I)\}\rangle&=& \frac{1}{\cal Z}\exp\bigg[\sum_I\epsilon_0(\lambda(I))\bigg]|\{\lambda(I)\}\rangle\cr\cr
&=& \frac{1}{\cal Z}\exp\bigg[\int dI \frac{\epsilon_0(\lambda(I))}{1+e^{\epsilon (\lambda (I))}} \bigg]|\{\lambda(I)\}\rangle\cr\cr
&=& \frac{1}{\cal Z}\exp\bigg[\int dI \langle \hat n_I \rangle \epsilon_0(\lambda(I)) \bigg]|\{\lambda(I)\}\rangle .\cr &
\end{eqnarray}
And so we see that Lagrange multiplier associated with the occupation number operator $\hat n_I$ is $\epsilon (\lambda (I))$.

While our view of the GGE differs from Ref. \cite{Ilievski2017} with its emphasis on $\rho_p(\lambda)$ as the fundamental object, it also differs 
from one where a microcanonical viewpoint is adopted \cite{   ,deNardis2017}.    In the microcanonical viewpoint one often
invokes the generalized eigenstate thermalization (gETH) hypothesis.  This hypothesis argues that one can employ a representative quantum state, $|s_\text{rep}\rangle$, in lieu
of performing a trace over a density matrix in computing the expectation value of any reasonable observable {\cal O}, i.e.
\begin{equation}
\langle s_\text{rep}|{\cal O}|s_\text{rep} \rangle = {\rm Tr}\hat\rho_\text{GGE} {\cal O}.
\end{equation}
In this viewpoint what is important is simply finding a representative state $|s_\text{rep}\rangle$.  By the gETH, any state that is characterized by an occupation
number of excitations given by 
\begin{equation}
\frac{\rho_p(\lambda) }{\rho_p(\lambda)+\rho_h(\lambda)} = \frac{1}{1+e^{\epsilon(\lambda)}}
\end{equation}
is equally good.  And so we see that in this picture it is $\epsilon(\lambda)$ (and not $\epsilon_0(\lambda)$) that becomes the primary quantity of interest.
Putting aside specific instances where the gETH is known to fail \cite{1742-5468-2014-9-P09026,Goldstein2014}, our interest in finding quasi-local bases of charges for quenches mandates that
we follow an approach to quantum quenches using a canonical density matrix.
\vskip 10pt
\hskip 10pt

\begin{acknowledgments}
We thank Neil Robinson, Enej Ilievski, and Milosz Panfil for valuable discussions. 
This research was funded by the U.S. Department of Energy under Contract No. DE-SC0012704.
\end{acknowledgments}

\appendix
\section{}
In this appendix we show how to arrive at Eqn. \ref{realspace}
demonstrating that the charges we are constructing are quasi-local at
$1/c$.  The spatial dependence of the charges is encoded in $F(x_1,x_2,x_3,x_4)$, defined as:
\begin{eqnarray}
F(x_1,x_2,x_3,x_4) &=& \cr\cr
&& \hskip -1.4in \int\frac{d\lambda_1}{2\pi}\int\frac{d\lambda_2}{2\pi}\int\frac{d\lambda_3}{2\pi}
C_{\lambda_1\lambda_2\lambda_3}e^{i\lambda_1x_{41}}e^{i\lambda_2x_{42}}e^{i\lambda_3x_{34}}.
\end{eqnarray}
To evaluate this, we first rewrite $C_{\lambda_1\lambda_2\lambda_3}$ as
\begin{eqnarray}
C_{\lambda_1\lambda_2\lambda_3} &=& \frac{1}{2}\left[\frac{\lambda_2-\lambda_3}{\lambda_1-\lambda_3}-\frac{\lambda_1-\lambda_3}{\lambda_2-\lambda_3}\right] \cr\cr
&&\hskip -.7in \times (q(\lambda_1)+q(\lambda_2)-q(\lambda_3)-q(\lambda_1+\lambda_2-\lambda_3)).
\end{eqnarray}
Performing a change of variables, $\{\lambda_1,\lambda_2,\lambda_3\}\to\{\alpha,\beta,\gamma\}$, with $\alpha=\lambda_1-\lambda_3$, $\beta=\lambda_2-\lambda_3$ and $\gamma$ set as the argument of $q$, we can easily evaluate this integral term by term,
\begin{equation}
\frac{1}{2}\int\frac{d\alpha}{2\pi}\int\frac{d\beta}{2\pi}\int\frac{d\gamma}{2\pi}\left(\frac{\beta}{\alpha}-\frac{\alpha}{\beta}\right)q(\gamma)\sum_{i=1}^4X_i(\alpha,\beta,\gamma).
\end{equation}
The exponents $X_i$ in the new variables read
\begin{align}
&X_1=e^{-i(\alpha x_{32}+\beta x_{24}+\gamma (x_1+x_2-x_3-x_4))};\\
&X_2=e^{-i(\alpha x_{14}+\beta x_{31}+\gamma (x_1+x_2-x_3-x_4))};\\
&X_3=-e^{-i(\alpha x_{14}+\beta x_{24}+\gamma (x_1+x_2-x_3-x_4))};\\
&X_4=-e^{-i(\alpha x_{32}+\beta x_{31}+\gamma (x_1+x_2-x_3-x_4))}.
\end{align}
Using 
\begin{equation}
\int\frac{d\alpha}{2\pi}\int\frac{d\beta}{2\pi}\frac{\alpha}{\beta}e^{-i\alpha x}e^{-i\beta y}=\frac{1}{2}\delta'(x)\text{sgn}(y),
\end{equation}
we then obtain our final expression for $F$:
\begin{eqnarray}
F(x_1,x_2,x_3,x_4)&=&\frac{1}{4}\tilde q(x_1+x_2-x_3-x_4)\cr\cr
&&\hskip -.8in \times \bigg[(\delta'(x_{24})\text{sgn}(x_{32})-\delta'(x_{32})\text{sgn}(x_{24}))\cr\cr
&& \hskip -.7in +(\delta'(x_{31})\text{sgn}(x_{14})-\delta'(x_{14})\text{sgn}(x_{31}))\cr\cr
&& \hskip -.7in -(\delta'(x_{24})\text{sgn}(x_{14})-\delta'(x_{14})\text{sgn}(x_{24}))\cr\cr
&& \hskip -.7in -(\delta'(x_{31})\text{sgn}(x_{32})-\delta'(x_{32})\text{sgn}(x_{31}))\bigg].
\end{eqnarray}

\bibliography{quasilocal}

\end{document}